\documentclass[11pt]{article}
\usepackage{latexsym}
\usepackage{amsmath}
\usepackage{amscd}
\usepackage{graphics}
\usepackage{psfig}
\graphicspath{{epsfigures/}}
\def\be#1\ee{\begin{equation}#1\end{equation}} 
\def\bea#1\eea{\begin{eqnarray}#1\end{eqnarray}} 

\newcommand{\nm}{\nonumber} 

\newcommand{\At}{{\widehat{A}}}
\newcommand{\mt}{{\widehat{m}}}
\newcommand{\xt}{{\widehat{x}}}
\newcommand{\yt}{{\widehat{y}}}
\newcommand{\Gt}{{\widehat{G}}}
\newcommand{\xit}{{\widehat{\xi}}}
\newcommand{\pt}{{\widehat{p}}}
\newcommand{\qt}{{\widehat{q}}}
\newcommand{\ft}{{\widehat{f}}}
\newcommand{\kt}{{\widehat{k}}}          

\newcommand{\Ab}{{\bar{A}}}
\newcommand{\mb}{{\bar{m}}}

\newcommand{\kb}{{\bar{k}}}

\newcommand{\blist}{\begin{list}{}{\setlength{\leftmargin}{4mm}
\setlength{\parindent}{0mm}\setlength{\parsep}{1mm}
\setlength{\topsep}{2mm}}}
\newcommand{\elist}{\end{list}}

\title{Many Accelerating Black Holes}

\author{ H.F. Dowker${}^{a, 1}$, 
S.N. Thambyahpillai${}^{b,2}$,\\
        $\;$ \\ ${}^1$ Dept. of Physics, 
        Queen Mary, University of London, London, UK.\\
        ${}^2$ Dept. of Physics, Harvard University,
        Cambridge MA, USA.\\
}
\begin{document}

\begin{titlepage}

\maketitle
\begin{abstract} 
\thispagestyle{empty} 

We show how the Weyl formalism allows metrics to be written down which
correspond to arbitrary numbers of collinear accelerating neutral black holes
in 3+1 dimensions.
The black holes have arbitrary masses and different accelerations and 
share a common acceleration horizon. In the general case, the black holes 
are joined by cosmic strings or struts that provide the necessary 
forces that, together with the inter black hole gravitational 
attractions, produce the acceleration. In the cases of two and three 
black holes, the parameters may be chosen so that the outermost
 black hole is pulled along by a cosmic string and the inner black holes
follow behind accelerated purely by gravitational forces. We conjecture that 
similar solutions exist for any number of black holes.

\end{abstract}
\vspace{0.1 cm}
\noindent ${}^a$f.dowker$@$qmw.ac.uk, 
${}^b$thamby$@$physics.harvard.edu.

\vspace{1cm}
\noindent
{QMUL-PH-01-05}, 
{HUTP-00/A039}

\end{titlepage}

\section{Introduction}

The Weyl formalism for static axisymmetric vacuum solutions in 3+1 
dimensions allows one to write down a solution that represents
an arbitrary number of collinear neutral black holes held
in unstable equilibrium by a system of struts and strings \cite{Israel:1964}. 
The metric is constructed using the Newtonian potential of 
a number of collinear rods of differing lengths and the same mass per unit 
length. 
The
thermodynamics of such configurations  has
been studied recently in \cite{Costa:2000kf}.
The Weyl construction
was used  by Myers to produce an infinite periodic 
array of neutral black holes, in which case the strings and struts disappear
\cite{Myers:1987rx}. 
It has also been long known that the C-metric 
for an accelerating black hole has an 
interesting Weyl form \cite{Godfrey:1972, Gibbons:1974,
Gibbons:1980nf, Bonnor:1983}. 
It is constructed from the Newtonian potential of 
a semi-infinite-line-mass (silm) and a second, finite, rod. The silm 
is the position of the acceleration horizon and the finite rod is the 
position of the black 
hole horizon. There is necessarily a cosmic string or strut 
in the spacetime which provides the force accelerating the 
black hole. It is now easy to guess the form of the metric which 
describes a number, $n$, of collinear black holes all accelerating  
in the same direction. It is constructed from the Newtonian 
potential of a silm and $n$ finite non-overlapping collinear rods of arbitrary 
lengths.  

We examine the $n=2$ case in detail. We show that the interpretation as
two accelerating black holes is valid. In the ``Newtonian limit'' where 
the accelerations are small compared to the inverse masses of the 
black holes, Newton's law holds for each black hole
with the acceleration produced by a 
combination of the forces due to any deficit angle cosmic strings or
struts present and the gravitational attraction of the other black hole.  

We show that the parameters may be chosen so that the outer
 black hole is pulled along by a cosmic string and the inner black hole
follows behind accelerated purely by its gravitational 
attraction to the outer one. A similar solution can be found for 
the $n=3$ case, in which the only deficit angle along the axis is 
a cosmic string attached to the outermost black hole and the 
other two are accelerated by gravitational forces alone. 
We conjecture that
similar solutions exist for any number of black holes.

\section{The Weyl form of the C-metric}

Any static axisymmetric vacuum solution of the Einstein equations
may be written in the following form
\be
\label{weylone.eq}
ds^2 = \alpha^2 (dr^2 + dz^2) + r^2 \gamma^{-2} d\phi^2 - \gamma^2
dt^2                                                                 
\ee
where $\alpha$,$\gamma$ are functions of $r$,$z$. By using the
substitutions
\begin{eqnarray}
\label{weyltwo.eq}
\alpha &=& e^{\nu-\lambda}  \nm \\
\gamma &=& e^{\lambda}
\end{eqnarray}
the above metric is transformed into the Weyl canonical form
\begin{equation}
\label{weylthree.eq}
ds^2 = e^{2(\nu - \lambda)} (dr^2 + dz^2) + r^2 e^{-2\lambda} d\phi^2
- e^{2\lambda} dt^2\ .
\end{equation}                  
 $\lambda$ must satisfy Laplace's equation
\begin{equation}
\label{weylfour.eq}
\Delta^{2} \lambda = {d^2\lambda\over dr^2} + {1\over r}{d\lambda\over
dr} + {d^2\lambda\over dz^2} = 0
\end{equation}
and $\nu$ can be found by using
\begin{eqnarray}
\label{weylfive.eq}
{d\nu\over dr} &=& r[{({d\lambda\over dr})^2} - {({d\lambda\over
dz})^2}]  \nm \\
{d\nu\over dz} &=& 2r {d\lambda\over dr}{d\lambda\over dz}\ .
\end{eqnarray}
                   
In \cite{Israel:1964} the case of a collinear array of arbitrarily spaced
black holes of arbitrary masses was studied. 
In that case the Newtonian potential used for $\lambda$ is the sum
of the potentials of a number of finite nonoverlapping rods placed along 
the $z$-axis and the function $\nu$ is found by quadrature.
Explicitly, let there be $n$ rods each centred at $z=a_i$, $i = 1,\dots n$, 
of length $b_i$ and with mass/unit length ${1\over2}$. 
The distances to the two ends of rod $i$ are given by 
\begin{eqnarray}
\label{weylsix.eq}
\rho_i^2 &=& r^2 + z_i^2  \nm \\
{\rho_{i}'}^2 &=& r^2 + {z_{i}'}^2  \nm \\
z_i &=& z - (a_i + {1\over 2}b_i)  \nm \\
z_{i}' &=& z - (a_i - {1\over 2}b_i)\ .
\end{eqnarray}                 

Then 
\be\lambda = \sum_{i = 1}^{n} \lambda_i(r,z)\ ;\quad 
    \nu = \sum_{i,j = 1}^{n} \nu_{ij} (r,z) \ ,\nm
\ee
where
\bea\lambda_i &=& {1\over 2} \ln \left({\rho_i' - z_i'\over \rho_i
-z_i}\right) \nm\\
\nu_{ij}&=& {1\over 4}\ln({E(i',j)E(i,j')\over E(i,j)E(i',j')})
\eea
where, for instance
\begin{equation}
\label{weyl eleven.eq}
E(i',j) = \rho_{i}'\rho_j + z_{i}'z_j + r^2 \ .\nm
\end{equation}                  

As stated in the introduction, the C-metric has 
the Weyl form built from the potential of a semi-infinite line mass
and a finite rod along the $z$-axis. 
 Consider the silm to stretch from a point
$z = c_0$ to $z=+ \infty$ and the finite rod to be between $z = c_2$ and
$z = c_1$ so that $c_2<c_1<c_0$. They both have mass/unit length
${1\over 2}$ as before. We are clearly free to choose the 
origin of $z$  coordinates to be wherever we like and 
so we fix it by requiring that the following three equations have
a solution for $A$ and $m$
\be\label{Aandm.eq}
2 A c_i^3 - c_i^2 + m^2 = 0, \quad i = 0,1,2 \ . \nm
\ee
In particular this means that $c_2< 0 < c_1 < c_0 $.

Then the following metric is the Weyl 
metric constructed from the corresponding Newtonian potential 
\bea\label{weylC.eq}
ds^2 =&-& A^{-1} X_0 X_1^{-1} X_2 dt^2 + 
A^{-1} r^2 [X_0 X_1^{-1} X_2]^{-1} d\phi^2 \nm\\
     &+& {m^2\over 4 A^3}(c_0-c_1)^{-2}(c_1-c_2)^{-2} 
{Y_{01}Y_{21}\over R_0 R_1 R_2 Y_{02}} (dz^2 + dr^2)
\eea
where
\bea
\zeta_i& =& z - c_i  \nm\\
R_i &=& (r^2 + \zeta_i^2)^{1\over 2}  \nm\\
X_i & =& R_i - \zeta_i  \nm\\
Y_{ij} & =& R_iR_j + \zeta_i\zeta_j + r^2\ . \nm
\eea

The overall scale and 
normalisations of the $t$ and $\phi$ coordinates are chosen with
hindsight so that the following coordinate transformation gives the 
C-metric in its most familiar form. Consider the coordinates $(x,y)$
related to $(z,r)$ by \cite{Godfrey:1972,
Bonnor:1983}
\bea
z &= & A^{-1} (x-y)^{-2} (1 - mAxy(x+y) - xy) \nm \\
r &= & A^{-1} (x-y)^{-2} (-G(y))^{1\over 2} (G(x))^{1\over 2}
\eea
where the function $G$ is the cubic 
\bea
G(\xi) &=& 1 - \xi^2 - 2mA \xi^3 \nm \\
       &= & - 2 m A (\xi-\xi_2)(\xi-\xi_3)(\xi-\xi_4)\ .
\eea
The roots of $G$, $\xi_2<\xi_3<\xi_4$
 are so labelled for historical reasons and there is the following 
relationship between them and the $c_i$
\be c_0= -m\xi_2, \quad c_1 = -m \xi_3, \quad c_2 = -m \xi_4 \ .
\ee

In terms of $(x,y)$ coordinates the metric (\ref{weylC.eq}) becomes the
familiar
\be\label{C.eq}
ds^2 = {1\over A^2(x-y)^2} \left[G(y)dt^2 - {dy^2\over G(y)} + 
G(x) d\phi^2 + {dx^2\over G(x)} \right]\ .
\ee
\section{Two accelerating black holes} 

In the maximally extended form of the ordinary C-metric above there are
actually two accelerating black holes, one covered by the coordinates
shown above and the other on the opposite side of the acceleration 
horizon, accelerating in the opposite direction and
out of causal contact with the first. It is possible however to construct
metrics in which there are two (or more) black holes on the 
same side of  a common acceleration horizon, accelerating in the 
same direction. (These then each have their ``mirror black hole''
on the other side of the acceleration horizon but we will be ignoring these
in what follows.) These are formed by taking the Weyl form with a 
potential constructed from a silm and two (or more) finite rods 
along the z-axis.
  
\begin{figure}[ht]
\centering
\resizebox{!}{1in}
{\includegraphics{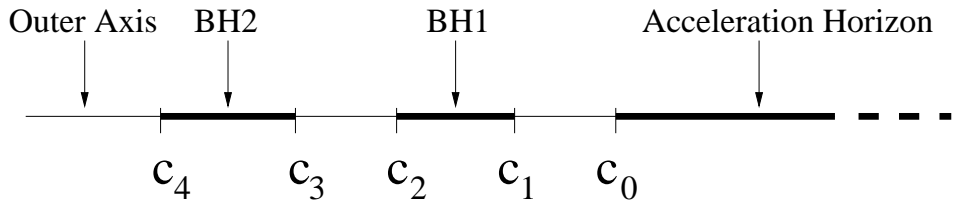}}
\caption{{
The thick lines correspond to the positions of line masses along
the $z$-axis and the dotted thick line indicates the infiniteness
of the righthandmost line mass.}
\label{fig1.fig}}
\end{figure}

Consider the silm and first rod as in the previous section and 
add a second rod between $z = c_4$ and $z= c_3$ where 
$c_4<c_3<c_2$ as illustrated in figure \ref{fig1.fig}. 

The corresponding Weyl metric is
\bea\label{weyl2.eq}
ds^2 =& -&\alpha^{-1} X_0 X_1^{-1} X_2 X_3^{-1} X_4 dt^2 + 
\alpha^{-1} r^2 [X_0 X_1^{-1} X_2 X_3^{-1} X_4]^{-1} d\phi^2 \nm\\
&+ &{\beta^2\over 4 \alpha}
{Y_{01}Y_{21}Y_{03}Y_{34}Y_{14}Y_{23}\over 2 R_0 R_1 R_2 
R_3 R_4 Y_{02} Y_{04} Y_{31} Y_{42}} (dz^2 + dr^2)
\eea
where $\alpha$ is a constant with units $L^{-1}$ and 
$\beta$ is a dimensionless constant which we discuss further below. 

Recall that we chose the origin of $z$ so that the equations
(\ref{Aandm.eq}) have a solution for $A$ and $m$. Now let $\kt$ 
be such that 
\be\label{Aandmhat.eq} 
2\At (c_i -\kt)^3 - (c_i -\kt)^2 + \mt^2 = 0, \quad i = 0,3,4 
\ee
have solutions for $\At$ and $\mt$. 

We want this metric to become the C-metric in three 
different limits
\bea (i) \quad c_1 & \rightarrow & c_2 \\
     (ii) \quad c_3 &\rightarrow & c_4 \\ 
     (iii) \quad c_2 &\rightarrow & c_3
\eea

In limit (ii) we want to obtain the C-metric 
exactly as it appears in (\ref{weylC.eq}). In limit ($i$) (($iii$)) 
the C-metric would be expressed in terms of 
parameters $\{c_0, c_3, c_4\}$ ($\{c_0,c_1,c_4\}$)  instead 
of $\{c_0, c_1, c_2\}$. 

In order for the limits to work we require $\alpha$ and $\beta$ to 
behave in the following ways
\bea (i) \quad (\alpha, \beta) &\rightarrow& 
         \left(\At, {\mt\over\At}(c_0-c_3)^{-1}(c_3-c_4)^{-1}
         \right)\quad \rm{as}\quad \ c_1 \rightarrow c_2 \\
     (ii) \quad (\alpha, \beta) &\rightarrow&
         \left(A, {m\over A}(c_0-c_1)^{-1}(c_1-c_2)^{-1}                        \right) \quad \rm{as}\quad \ c_3 \rightarrow c_4 \\
     (iii) \quad (\alpha, \beta) &\rightarrow&
         \left(\Ab, {\mb\over\Ab}(c_0-c_1)^{-1}(c_1-c_4)^{-1}                        \right) \quad \rm{as}\quad \ c_2 \rightarrow c_3
\eea      

where $\Ab$ and $\mb$ depend on $c_0, c_1, c_4$ via
\be 2\Ab(c_i - \kb)^3 - (c_i - \kb)^2 + \mb^2 =0, \quad i = 0,1,4
\ee
for some $\kb$. 

These conditions are not enough to fix the dependence of
$\alpha$ and $\beta$ on $c_0,\dots c_4$ but there are 
solutions. 

Let Black Hole 1 (BH1) coordinates $(x,y)$ be defined as before
\bea
z& = & A^{-1} (x-y)^{-2} (1 - mAxy(x+y) - xy) \nm \\
r& = & A^{-1} (x-y)^{-2} (-G(y))^{1\over 2} (G(x))^{1\over 2}\ .
\eea            
Let Black Hole 2 (BH2) coordinates, $(\xt, \yt)$ be given by 
\bea
z - \kt &= & \At^{-1} (\xt-\yt)^{-2} (1 - \mt\At\xt\yt(\xt+\yt) - \xt\yt) \nm \\
r &= & \At^{-1} (\xt-\yt)^{-2} (-\Gt(\yt))^{1\over 2} (\Gt(\xt))^{1\over 2}\ .  
 \eea
where the function $\Gt$ is the cubic
\bea
\Gt(\xit)&=& 1 - \xit^2 - 2\mt\At \xit^3 \nm \\
       &=& - 2 \mt \At (\xit-\xit_2)(\xit-\xit_3)(\xit-\xit_4)
\eea
and
\be c_0 -\kt = -\mt\xit_2, \quad c_3 -\kt= -\mt \xit_3, 
\quad c_4 -\kt= -\mt \xit_4 \ .
\ee                         
Then we can transform the metric (\ref{weyl2.eq}) into BH1 and BH2 coordinates
and it becomes, respectively
\bea\label{BH1.eq}
ds^2 = &&{1\over A^2 (x-y)^2}\Bigg[ {A\over \alpha}\left(
               {X_4\over X_3}G(y) dt^2 + {X_3\over X_4}G(x) d\phi^2\right)\\
& +& {\beta^2 A^3\over \alpha m^2}(c_0-c_1)^{2}(c_1-c_2)^{2}
    {Y_{03}Y_{34}Y_{14}Y_{23} \over 2 R_3 R_4 Y_{04} Y_{31} Y_{42}}
\left( -{dy^2\over G(y)} + {dx^2\over G(x)} \right) \Bigg]
\eea
and
\bea\label{BH2.eq}
ds^2 =&& {1\over \At^2 (\xt-\yt)^2}\Bigg[ {\At\over \alpha}\left(
               {X_2\over X_1}\Gt(\yt) dt^2 + 
{X_1\over X_2}\Gt(\xt) d\phi^2\right)\nm\\
& +& {\beta^2 \At^3\over \alpha \mt^2}(c_0-c_3)^{2}(c_3-c_4)^{2}
    {Y_{01}Y_{12}Y_{32}Y_{41} \over 2 R_1 R_2 Y_{02} Y_{13} Y_{24}}
\left( -{d\yt^2\over \Gt(\yt)} + {d\xt^2\over \Gt(\xt)} \right) \Bigg]\ .
\eea        
                            
Let us examine the solution close up to BH1 in BH1 coordinates. 
We consider the limit $r \rightarrow 0$ and $c_2< z < c_1$.
In that limit, metric (\ref{BH1.eq}) tends to 
\be\label{nearBH1.eq}
{A^{-2}\over (x-\xi_2)^2} \left[f(x)\left(p(y-\xi_2)G'(\xi_2)dt^2 - 
q{dy^2\over(y-\xi_2) G'(\xi_2)} + q {dx^2\over G(x)}
\right) + p
{G(x)\over f(x)} d\phi^2 \right]
\ee          
where
\bea
f(x)& =& {A^{-1}(x-\xi_2)^{-2}(1 - mAx\xi_2(x+\xi_2) - \xi_2 x) - c_3
\over A^{-1}(x-\xi_2)^{-2}(1 - mAx\xi_2(x+\xi_2) - \xi_2 x) -c_4}\nm\\
p & =& A \alpha^{-1}\nm\\
q & =& {\beta^2\over \alpha}{A^3\over m^2}(c_1-c_4)^2(c_0-c_1)^2
(c_1-c_2)^2(c_0-c_3)^2
(c_1-c_3)^{-2}(c_0-c_4)^{-2}\ .
\eea

From this we cannot read off the surface gravity of the black hole
as we can in the asymptotically flat case because the timelike coordinate
$t$ is not Minkowski time at spatial infinity but, rather, a Rindler-like
time. To see this we consider the limit of (\ref{weyl2.eq}) as $r,z
\rightarrow\infty$. 
\be
ds^2 \rightarrow -\zeta^2 d\eta^2 + d\zeta^2 + d\rho^2 + \rho^2 d\psi^2
\ee
where
\bea 
\zeta^2&=& {\beta^2\over \alpha} \left(\sqrt{r^2+z^2} - z\right)
\nm\\
\rho^2&=& {\beta^2\over \alpha} r^2 \left(\sqrt{r^2+z^2} - z\right)^{-1}
\nm\\
\eta& =& {1 \over \beta} t \nm\\
\psi& =& {1 \over \beta} \phi
\eea
and $\beta = (c_0-c_1)(c_1-c_2)(c_0-c_3)(c_3-c_4)$.
This tells us that the period, $\Delta\tau_{\rm{AH}}$,
 of the euclidean time coordinate $\tau = it$ that would be 
required for the euclideanized solution to have no conical singularity 
at the acceleration horizon is $\Delta\tau_{\rm{AH}} = 2 \pi 
\beta$.

We can, however, read off from (\ref{nearBH1.eq})
the period, $\Delta\tau_{\rm{BH1}}$,
 that the euclidean time coordinate
$\tau = i t$ would have to have in order for the euclideanized solution 
to be regular on the horizon of BH1. It is
\be \Delta\tau_{\rm{BH1}} = 2 \pi {\beta}
{(c_1-c_2)(c_0-c_3)(c_1-c_4) 
\over(c_0-c_2)(c_0-c_4)(c_1-c_3)} 
\ee
 
We can also examine the smoothness conditions along the axis in 
either direction from BH1. Let the period of $\phi$ necessary for
smoothness on the axis pointing towards the acceleration horizon, 
{\it i.e.} $r=0$ and $z=c_1$, be $\Delta\phi_{\rm{1R}}$ and
the period of $\phi$ necessary for
smoothness on the axis pointing towards the outer black hole BH2, 
{\it i.e.} $r=0$, $z=c_2$ be $\Delta\phi_{\rm{1L}}$. 
When $r$ is small, the limit $z \rightarrow c_1$ corresponds to 
$x\rightarrow \xi_4$ and 
examining (\ref{nearBH1.eq}) in that limit we can read off
\be \Delta\phi_{\rm{1R}} = 2 \pi {\beta}
{(c_0-c_1)(c_0-c_3) \over(c_0-c_2)(c_0-c_4)}
\ee               
On the other side, $z \rightarrow c_2$ when $x\rightarrow \xi_3$
and
\be \Delta\phi_{\rm{1L}} = 2 \pi {\beta}
{(c_1-c_4)(c_2-c_3) (c_0-c_3)\over(c_1-c_3)(c_0-c_4)
(c_2-c_4)}
\ee   

We can repeat this all for BH2, looking at  
(\ref{BH2.eq}) near the horizon of BH2.
In this limit the metric becomes
\be\label{nearBH2.eq}
{{\At}^{-2}\over (\xt-\xit_2)^2} \left[\ft(\xt)
\left(\pt(\yt-\xit_2){\Gt}'(\xit_2)dt^2 
- \qt{d\yt^2\over (\yt-\xit_2){\Gt}'(\xit_2)}
+ \qt{d\xt^2\over \Gt(\xt)}
\right) + \pt
{\Gt(\xt)\over \ft(\xt)} d\phi^2 \right]
\ee
where
\bea
\ft(\xt)& =& {\At^{-1}(\xt-\xit_2)^{-2}(1 - \mt\At\xt\xit_2(\xt+\xit_2) - \xit_2 \xt) - c_2
\over \At^{-1}(\xt-\xit_2)^{-2}(1 - \mt\At\xt\xit_2(\xt+\xit_2) - \xit_2 \xt) -c_1}\nm\\
\pt & =& \At \alpha^{-1}\nm\\
\qt & =& {\beta^2\over \alpha}{\At^3\over \mt^2}(c_1-c_4)^2(c_0-c_3)^{2}
(c_3-c_4)^{2}(c_2-c_4)^{-2}
\eea                                 

The period of $\tau=it$ required for regularity on the BH2 horizon is 
\be \Delta\tau_{\rm{BH2}} = 2 \pi \beta 
{(c_1-c_4)(c_3-c_4) \over(c_2-c_4)(c_0-c_4)}
\ee           
The period of $\phi$ required for smoothness on the axis pointing towards
BH1 and the acceleration horizon is 
\be 
\Delta\phi_{\rm{2R}} = 2 \pi {\beta}
{(c_1-c_4)(c_2-c_3)(c_0-c_3) \over(c_1-c_3)(c_0-c_4)
(c_2-c_4)}
\ee         
Notice that this is equal to $\Delta\phi_{\rm{1L}}$ as it should be 
because these refer to the same part of the axis, that 
 between the black holes. 
The period of $\phi$ required for smoothness on the outer axis 
is
\be \Delta\phi_{\rm{2L}} = 2 \pi {\beta}\ .
\ee                           

It is possible to choose the parameters $c_i$, $i = 0,\dots 4$ so that
$\Delta\tau_{\rm{BH1}}= \Delta\tau_{\rm{BH2}}$.
Indeed this condition implies
\be (c_0-c_4)(c_1-c_2) = (c_0-c_2)(c_3-c_4)
\ee
which can be rewritten as
\be (c_1-c_2)\left[(c_0-c_3) +(c_3 -c_4)\right]
= \left[(c_0-c_1) + (c_1-c_2)\right](c_3-c_4)
\ee
and finally as
\be
(c_3 - c_4) = (c_1-c_2)(c_0-c_3)
              (c_0-c_1)^{-1}\ .
\ee
Any choice of $c_0, c_1, c_2$ and $c_3$ can be made and $c_4$ 
is then fixed.

 But it is not possible to have  
either $\Delta\tau_{\rm{BH1}}= \Delta\tau_{\rm{AH}}$
or $\Delta\tau_{\rm{BH2}}= \Delta\tau_{\rm{AH}}$, 
which is not a surprise given the same is true for the 
C-metric.
Thus, on euclideanizing the solution we may choose the period
of imaginary time to make both black hole
horizons regular but if we do so there will be 
a conical singularity at the acceleration horizon.

We check that in the weak field limit, {\it i.e.} $mA << 1$ and
$\mt\At << 1$ and when the distance between the black holes is
large compared to their masses, the acceleration of BH2 is 
given by Newton's laws. We may realise this limit by taking 
$c_0 \rightarrow \infty $ and keeping  
 $c_1-c_2$ and $c_3-c_4$ fixed. 
In this limit we have the following 
\bea
c_0-c_1 &\rightarrow& {1\over 2A} - m + O(mA) \\
c_1-c_2 &\rightarrow& 2 m + O((mA)^2) \\
c_0-c_3 &\rightarrow& {1\over 2\At} - \mt + O(\mt\At) \\
c_3-c_4 &\rightarrow& 2\mt + O((\mt\At)^2)\ .
\eea
The net force outwards (to the left) on BH2 
due to the combination of possible 
strings and struts is given by 
\be
T_2 = {1\over 4}\left[{\Delta\phi_{\rm{2L}}\over \Delta\phi_{\rm{2R}}} -1
              \right]
\ee
Expanding this out and setting $c_2-c_3 = R$ we find
\be\label{newtontwo.eq}
T_2 = \mt \At + {m\mt\over R^2} + \rm{correction} 
\ee
The first terms of the correction are cubic in ${m\over R}$ and 
${\mt\over R}$ and quadratic in $mA$ and  $\mt \At$. In order for the 
above to be a consistent approximation we require that $R$ not be
fixed but tends to infinity 
in such a way that ${m\over R}$ and ${\mt\over R}$ are larger than 
$mA$ and $\mt\At$ in the limit. We might choose, 
for example, $c_0 \rightarrow \infty$,
$c_1-c_2$ and $c_3-c_4$ fixed and $R \sim \sqrt {c_0}$.  

Similarly the net force to the left (towards BH2) on BH1 due
to the strings and/or struts is 
\be
T_1 = {1\over 4}\left[{\Delta\phi_{\rm{1L}}\over \Delta\phi_{\rm{1R}}} -1
              \right]
\ee                   
and in the same limit as above this gives
\be\label{newtonone.eq}
T_1 = m A - {m\mt\over R^2} + \rm{correction}
\ee    

Then (\ref{newtontwo.eq}) and (\ref{newtonone.eq})
 show that the acceleration of each black hole is 
due to the net tension on it due to the strings and/or struts
together with the gravitational attraction of 
the other black hole.

We can ask, is it possible to choose the period of $\phi$ so that 
BH1 is free of strings or struts so that its acceleration is due
entirely to its gravitational attraction towards BH2? 
The condition for this is  $\Delta \phi_{\rm{1L}}
= \Delta\phi_{\rm{1R}}$. This reduces to the condition
\be\label{nostrings.eq}
 (c_0-c_1)(c_3-c_4) = (c_1-c_4)(c_2-c_3) \ .
\ee
This can be satisfied by choosing $(c_3-c_4)$, $(c_2-c_3)$ and
$(c_1-c_4)$ arbitrarily. $(c_0-c_1)$  is independent and 
can be chosen to satisfy (\ref{nostrings.eq}). 
Note that in the weak field limit as defined above this condition
is approximately
$ A = \mt/R^2$ so indeed the acceleration of the black hole is 
due to its gravitational attraction towards the outer one.

\section{More than two black holes}

It is clear how to generalize the above results to the case of
three or more black holes. In the case of three, 
we add a third finite rod on the $z$-axis 
between $z=c_6$ and $z=c_5$ to
the left of the existing two. The metric will be constructed from 
(\ref{weyl2.eq}) by multiplying $g_{tt}$ by $X_6/X_5$ and 
$g_{\phi\phi}$ by the inverse of this. The rest of the 
metric is multiplied by a factor
\be
{Y_{05}Y_{56}Y_{16}Y_{52}Y_{36}Y_{45} \over 2R_5R_6 Y_{06} Y_{15}
Y_{26} Y_{35} Y_{46}}\ .
\ee
$\beta$ and $\alpha$ will now be functions of the parameters $c_0,
\dots c_6$ such that the metric reduces to the known C-metric in the 
appropriate limits. Adding more black holes proceeds similarly.

We can again calculate the various periods required for regularity.
We can also ask whether the $c_i$ may be chosen so that there is
only one cosmic string pulling the outermost black hole with the
others being accelerated purely because of the gravitational 
attraction of the others. This is a more difficult question to
answer in the general 
case. In the case of three black holes there are two conditions,
\bea
& c_{14} c_{23} c_{16} c_{25} c_{02} = c_{01} c_{13} c_{24} c_{15} c_{26}\\
& c_{16} c_{25} c_{36} c_{45} c_{02} c_{04} = 
     c_{01} c_{03} c_{15} c_{26} c_{35} c_{46}
\eea
where $c_{ij} = c_i - c_j$. These do have solutions with $c_i < c_j$ for
$i>j$, for example: 
\be
c_6 = 0,\quad c_5= a, \quad c_4=a+1, \quad c_3=a+2, \quad c_2=a+3,
\quad c_1=a+4, \quad c_0 = a + 4 + f
\ee
where $f = (9a -6)/19$,  and $a$ is the positive
root of $3z^2 - 5z - 36$. We conjecture that such solutions do exist 
for any number of black holes. Then one could conceive of an 
accelerating version of the Myers black hole array in which there is 
an infinite collinear array of accelerating black holes with a 
common acceleration horizon and no cosmic strings or struts in
sight. We also conjecture that there are solutions with any number 
of accelerating black holes 
for which it would be possible to 
choose the period of imaginary time to be such that the 
euclidean solution is regular on all the black hole horizons at once.   

\section{Discussion}

These multi accelerating black 
hole solutions are classically unstable, 
assuming that the deficit  angles of the 
nodal singularities are fixed as they would be if
the solutions are 
approximations to spacetimes with field theory 
cosmic strings of fixed tension. A little displacement of one of the
black holes will disturb the 
balance of forces required to maintain the solution and 
presumably cause some black holes  
to collide and/or some to be left behind. 
This instability prevents us from 
asking whether the euclideanized solutions can be interpreted as 
instantons for multi-pair production of black holes or the 
production of some pairs in the background of other accelerating 
pairs. This
is even before we confront the problem that
the uncharged solutions do not give rise to  regular 
instantons --- even if we allow cosmic strings in space --- because 
of the impossibility of matching the acceleration and black hole
horizon temperatures. (Nevertheless the euclidean action may 
straightforwardly be calculated as in \cite{Eardley:1995au} 
since the singularities are conical and integrable. Using the 
work in \cite{Eardley:1995au}, calculating the action is just a matter
or calculating the difference in the acceleration horizon areas
between the instanton and background and the areas of the horizons
of the black holes produced.)

These problems might be resolved were we able to give charge to the 
black holes. The obvious method to try is that of the charging 
transformation given in \cite{Costa:2000kf}. However this only works when the
time coordinate is asymptotically Minkowki time (and
in any case gives all the black holes the same charge 
to mass ratio and so might not be able to produce a regular instanton).
Here, our asymptotic time 
coordinate is Rindler time and the charging 
transformation gives a new metric of unknown 
physical interpretation which has zero net charge
at infinity. 
Finding charged versions of these 
solutions requires more ingenuity. 
 
\section{Acknowledgments} We would like to thank Jerome 
Gauntlett, David Kastor,  Rob Myers and Harvey Reall for useful 
discussions and Gary Gibbons for 
telling us about the Weyl form of the C-metric. 
FD was supported in part by an EPSRC Advanced Fellowship.

\providecommand{\href}[2]{#2}\begingroup\raggedright\endgroup
% \bibliography{refs}
% \bibliographystyle{utphys}
\end{document}